# DERIVATIVE-BASED MIR SPECTROSCOPY FOR BLOOD GLUCOSE ESTIMATION USING PCA-DRIVEN REGRESSION MODELS

Saeed Mansourlakouraj
School of Electrical Engineering
Sharif Univeristy of Technology

Hadi Barati
School of Electrical Engineering
Sharif Univeristy of Technology

Mehdi Fardmanesh
School of Electrical Engineering
Sharif Univeristy of Technology

December 10, 2024

ABSTRACT

In this study, we presented two innovative methods, which are Threshold-Based Derivative (TBD) and Adaptive Derivative Peak Detection(ADPD), that enhance the accuracy of Learning models for blood glucose estimation using Mid-Infrared (MIR) spectroscopy. In these presented methods, we have enhanced the model's accuracy by integrating absorbance data and its differentiation with critical points. Blood samples were characterized with Fourier Transform Infrared (FTIR) spectroscopy and advanced preprocessing steps. The learning models were Ridge Regression and Support Vector Regression(SVR) using Leave-One-out Cross-Validation. Results exhibited that TBD and ADPD significantly outperform basic used methods. For SVR, the TBD increased the $r^2$ score by around 27%, and ADPD increased it by around 10%; these Ridge Regression values were between 36% and 24%. In addition, Results demonstrate that TBD and ADPD significantly outperform conventional methods, achieving lower error rates and improved clinical accuracy, validated through Clarke and Parkes Error Grid Analysis.

*K*eywords Blood Glucose Estimation, Mid-Infrared Spectroscopy, Fourier Transform Infrared Spectroscopy (FTIR), Derivative Spectroscopy, Threshold-Based Feature Selection, Support Vector Regression (SVR), Ridge Regression

## 1 Introduction

Two different types of infrared spectroscopy are utilized for blood glucose measurements. Mid-infrared (MIR) technique
([1] and [2]) measures the absorption of infrared light in the frequency range of 2.5 – 25$\mu m$ (400 – 4000$cm^{-1}$) while in the near-infrared (NIR) approach, the absorption is detected in the range of 700$nm$ – 2.5$\mu m$ (4000 – 14000$cm^{-1}$) [1]. The blood glucose (BG) level can be estimated by detecting the glucose molecules in the blood utilizing Fourier Transform Infrared Spectroscopy (FTIR). This method can be seen as a non-invasive alternative to traditional fingerprick glucometers, which measure blood glucose in an invasive and rather painful approach. The main weakness of this approach is the weak infrared absorption of glucose, resulting in essential data preprocessing and chemometric analysis [3]. Various multivariate calibration methods such as partial least-squares regression (PLSR), principal component regression (PCR), and artificial neural network (ANN) can be utilized to quantify the blood glucose level in blood samples collected from subjects, such as patients in a general hospital or healthy individuals. Comparing the standard error of prediction of various methods showed that the combined PLSR-ANN method generated smaller errors compared to those produced by PLS or PCR methods employed individually [4]. The non-linear relationship between spectral data and blood glucose concentration leads to limitations in utilizing linear calibration models, and thus hybrid models can alleviate these limitations. For example, a hybrid model based on integrated linear PLSR with the nonlinear stacked auto-encoder (SAE) deep neural network significantly optimizes the prediction of BG concentration from the measured diffuse reflectance spectrum of the palm compared to traditional PLSR model predictions [5]. Accurate prediction of the BG level can be achieved by employing various machine learning approaches such as Support Vector Machine Regression (SVMR), Random Forest Regression (RF), Extra Trees Regression (ETR), eXtreme Gradient Boosting (Xgboost), and hybrid Principal Component Analysis-Neural Network (PCA-NN). For instance, the measured absorbance spectra of a number of glucose aqueous samples were randomly split into 80% for the training set and 20% for the testing set. It was found that the models SVMR, ETR, and PCA-NN reached extremely good performances [6].

To improve the accuracy of BG level prediction in the MIR approach, the wavenumber importance index was generated by an effective variable selection method resulting in a reduction in time cost [2]. Investigating the correlation between the BG level and the measured MIR transmittance, it was seen that the measurement in the range of 3000 – 3500 $cm^{-1}$ was effective [7]. It has been shown that the combination of FTIR spectroscopy with univariate and multivariate chemometric analysis can effectively be treated as an innovative, non-invasive, and sustainable approach to distinguish nondiabetic subjects from diabetic and diabetic-insulin treated ones [8].

In this study, we are going to develop two methods, Threshold-based derivative (TBD) and Adaptive Derivative Peak Detection(ADPD), to enhance the prior model's accuracy; these models are based on combining the absorbance signal and its derivative. Ultimately, to demonstrate their strength and robustness, we will test them on two regression models: Support Vector Regression and Ridge Regression.

## 2 Materials and methods

### 2.1 Sample preparation

To capture a wide range of glucose values,50 samples were taken at different times from five healthy subjects aged 21 to 28 years old, including three men and two women. Following sample collection, blood glucose levels, which were in the range of 72 to 125 mg/dl, were determined using a standard glucometer, which also supplied the reference labels for the FTIR data. A Bruker Vertex 70 spectrometer operating in the mid-infrared (MIR) was utilized for sampling. To guarantee the accuracy and consistency of the data, four samples were also eliminated from the study since they did not fit the requirements for inclusion.

### 2.2 FTIR Device Configuration and Setup

A Bruker Vertex 70 spectrometer was used to perform the Fourier Transform Infrared (FTIR) measurements for this study. With a resolution of 1$cm^{-1}$, the spectrometer was set up to function precisely, which is necessary for catching spectrum information relevant to our research. The 400˘4000 1$cm^{-1}$ spectral wave number band was chosen since it corresponds to MIR. The samples were measured in transmittance, and in order to improve the signal-to-noise ratio, the spectrum was derived by averaging 32 scans.

The device had a KBr broadband beam splitter installed. Additionally, a background spectrum was captured before every sample measurement to guarantee the data's accuracy.

### 2.3 Pre-processing

The data should be pre-processed before applying innovative methods and machine learning models to ensure optimal performance. The given data was in transmittance format, which needs to be converted to absorbance format, using the below formula [9]:

$$A = -\log_{10}(T)$$

Where:
- **A** is the absorbance *i*,
- **T** is the transmittance

After that I applied rubber band correction, that is an important pre-processing step to remove any background offsets or slopes in the spectra which are measured by FTIR [10]. Additionally, I used min-max normalization using each sample to scale the data to a common range.

For smoothing the Savitzky-Golay algorithm [11] was applied, which is one of the most commonly used approach for spectroscopic data [12]. The chosen order was 2 with a window size of 100 to preserve the key spectral features.

## 2.4 Threshold-based derivative

We got the idea of combining absorbance and its first derivative from a study that demonstrated the benefit of applying derivatives in spectral treatments to improve signal differentiation, particularly for resolving overlapping spectral features [13]. In another study, the authors applied the first and second derivatives to Fourier Transform Infrared (FTIR) spectroscopy data, which decreased the error in multivariate models such as Partial Least Squares Regression (PLSR) and Principal Component Regression (PCR) by enhancing sensitivity and precision in detecting and quantifying components [14].

However, while derivatives enhance differentiation, they tend to remove key features like peaks, which are crucial in FTIR data. These peaks represent specific vibrational modes and carry significant information about the molecular composition of biological samples [15]. To overcome this limitation, we developed a threshold-based method that balances the benefits of both absorbance and its derivative. Specifically, for each sample, we calculate the first derivative of the absorbance spectrum, multiply it by 100 to address the low amplitude of the derivative, and then apply a threshold. If the scaled derivative at a particular wavenumber is below the threshold, we retain the original absorbance value; otherwise, we use the scaled derivative. Mathematically, this can be represented as follows:

$$F_i = \begin{cases} A_i & \text{if } \left|\frac{dA_i}{dw}\right| < \tau \\ \frac{dA_i}{dw} & \text{otherwise} \end{cases}$$

Where:

- $A_i$ is the absorbance at wavenumber *i*,
- $\frac{dA_i}{dw}$ is the derivative of absorbance with respect to wavenumber,
- $\tau$ is the threshold.

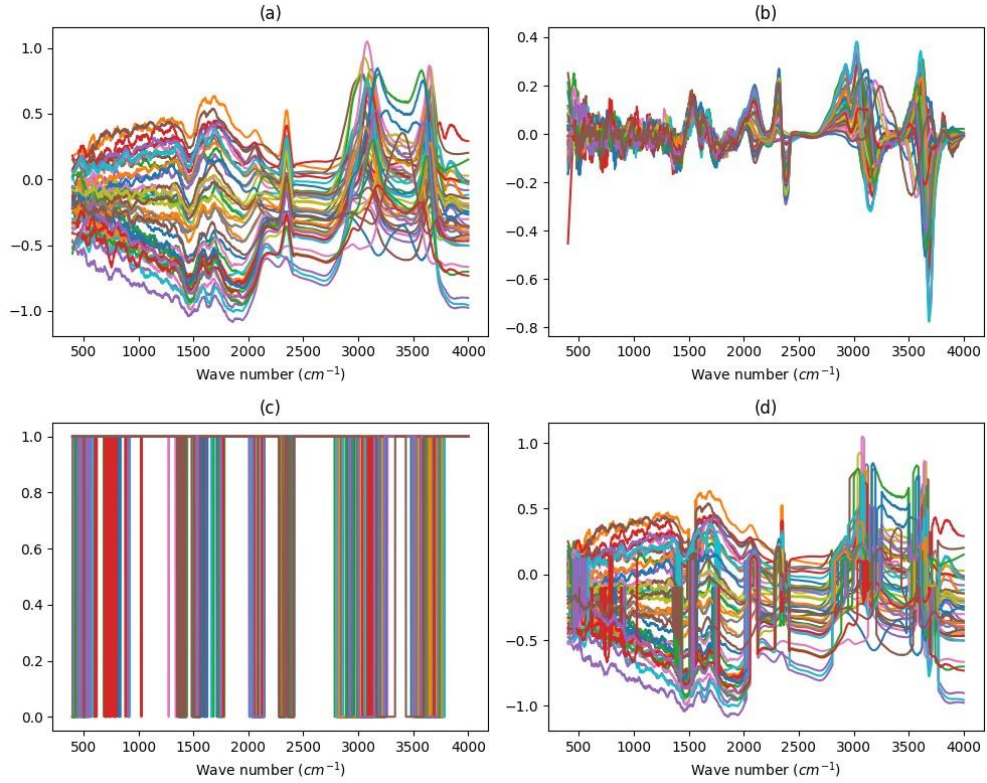

Figure 1: Illustration of the processing steps: a: The original absorbance spectrum. b: The first derivative of the absorbance spectrum, multiplied by 100 to increase its amplitude for improved analysis. c: The digitized threshold result, where a value of 11 indicates retention of the original absorbance value and a value of 1 indicates use of the derivative. d: The final processed result combining absorbance and derivative data, balancing peak information with enhanced differentiation.

By applying this method, we preserve the critical peak information from the absorbance spectrum while also benefiting from the enhanced differentiation provided by the derivative. Our results show that this approach improves the model's performance compared to using only the derivative or absorbance. The whole process could be seen in Figure 2.

2.5   Adaptive Derivative Peak Detection

Also another dynamic approach is developed to enhance the estimation of blood glucose levels. The method integrates both raw absorbance data, denoted as *x*, and its derivative, denoted as *z*, to improve spectral feature extraction. Since spectral peaks are crucial for accurate glucose level estimation [16], a dynamic relationship between *x* and *z* is introduced through the equation:

$$y = x - \alpha z \cdot x$$

where *z* represents the first derivative of *x* with respect to wavenumber, and $\alpha$ is a dynamically adjusted parameter. This parameter $\alpha$ is optimized to ensure that peaks in the spectra, which carry important information [15], are emphasized while reducing the influence of noise from the derivative data. The dynamic adjustment of $\alpha$ allows the model to retain important spectral features by balancing the use of raw and derivative data.

By applying this method, we preserve the critical peak information from the absorbance spectrum while balancing the needed amplitude for being a peak. Our results show that this approach improves the model's performance compared to using only the derivative or absorbance. The whole process could be seen in Figure 2.

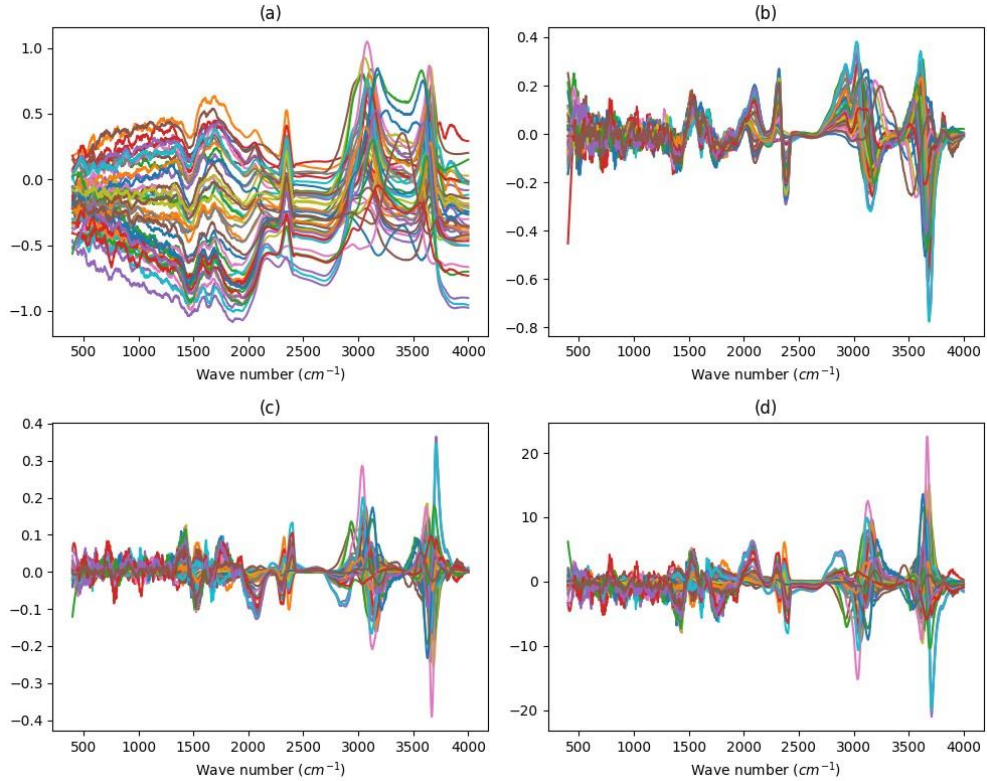

Figure 2: Illustration of the processing steps: a: The original absorbance spectrum. b: The first derivative of the absorbance spectrum, multiplied by 100 to increase its amplitude for improved analysis. c: Multiples of the first derivative and the absorbance signal d: The final processed result, which is used for learning

## 2.6 Model Training and Evaluation

To predict blood glucose levels from the feature vectors extracted from the MIR spectroscopy data, we built and evaluated several machine learning models, including *Ridge Regression*, *Support Vector Regression (SVR)*. Given the small size of our dataset, we employed *Leave-One-Out Cross-Validation (LOOCV)* to assess the performance of these models. LOOCV is suitable for small datasets as it maximizes the usage of all available data while helping to mitigate overfitting [17].

The models were evaluated using two standard error metrics: *Root Mean Square Error (MSE)* ([18]and [19]) and *Mean Absolute Error (MAE)* [20]. MSE emphasizes larger errors by squaring the residuals before averaging, while MAE provides an average of the absolute differences between the predicted and actual values.

Before applying the models, we normalized the feature vectors to ensure that all features contributed equally to the prediction task, as the features were originally on different scales. The normalization was performed using the training split of each fold during LOOCV. Specifically, we applied *standard normalization*, which transforms the data to have zero mean and unit variance [21].

$$z = \frac{x - \mu}{\sigma}$$

where:

- *x* is the original feature value,

- $\mu$ is the mean of the feature values in the training set,
- $\sigma$ is the standard deviation of the feature values in the training set,
- $z$ is the normalized feature value.

The mean $\mu$ and standard deviation $\sigma$ were calculated from the training set of each cross-validation fold, and the same parameters were then applied to normalize the test data in that fold. This ensures that no information from the valid data leaked into the training process.

## 2.7 Learning Models

In this study, we have used two well-known models, the ridge regression and SVR to demonstrate the validity of two proposed methods, ADPD and TBD.

### 2.7.1 Ridge Regression

Ridge regression (also known as Tikhonov regularization) [22] is a variant of linear regression that incorporates a regularization term to address multicollinearity and overfitting issues. It introduces a penalty to the sum of squared coefficients, which helps in reducing the coefficient values and stabilizing the model.

$$\hat{\beta}_{\text{ridge}} = \left(\mathbf{X}^T\mathbf{X} + \alpha \mathbf{I}\right)^{-1} \mathbf{X}^T \mathbf{y}$$

where:

- $\mathbf{X}$ is the matrix of input features,
- $\mathbf{y}$ is the response variable,
- $\alpha$ is the regularization parameter (also called the shrinkage parameter), and
- $\mathbf{I}$ is the identity matrix.

### 2.7.2 Support Vector Regression (SVR)

Support Vector Regression (SVR), originally introduced by Vapnik ([23] and [24]), is a method designed to handle both linear and nonlinear relationships between input variables (features) and the target variable. Additionally, the complexity of the model is controlled to prevent overfitting. The mathematical formulation [25] of the SVR can be expressed as $y_i = \mathbf{w}^T\mathbf{x}_i + b$ where:

$$\begin{aligned}
\underset{\mathbf{w},b,\xi_i,\xi_i^*}{\text{minimize}} \quad & \frac{1}{2}\|\mathbf{w}\|^2 + C\sum_{i=1}^{n}(\xi_i + \xi_i^*) \\
\text{subject to} \quad & y_i - \mathbf{w}^T\mathbf{x}_i - b \leq \epsilon + \xi_i \\
& \mathbf{w}^T\mathbf{x}_i + b - y_i \leq \epsilon + \xi_i^* \\
& \xi_i, \xi_i^* \geq 0, \quad \text{for } i = 1, \ldots, n
\end{aligned}$$

where:

- $\mathbf{w}$: Weight vector, representing the model's coefficients.
- $b$: Bias term, accounting for the offset in the model.
- $\xi_i, \xi_i^*$: Slack variables, allowing deviations beyond the $\epsilon$ margin.

- *C*: Regularization parameter, controlling the trade-off between model complexity and tolerance of deviations.
- *ϵ*: Insensitivity margin, within which predictions are not penalized for deviations.

For this study, we utilized a variety of kernels in the training phase, including linear, RBF, and polynomial kernels, to identify the best fit. The range of hyperparameters for this model is shown in Table 1.

## 3 Results and discussion

In each model, Principal Component Analysis (PCA) was utilized within the hyperparameter tuning process ([6] and [26]). The number of principal components was adjusted from 1 to 20, enabling models to identify the optimal number of components that balanced dimensionality reduction with accuracy(Here MSE). This approach allowed the models to capture the most relevant variance in the data while minimizing noise and reducing the likelihood of overfitting which is due to excessive complexity, leading models to focus too much on quirks in the sample data instead of identifying real, meaningful patterns.

The results of our study showed that the preprocessing method we developed significantly improved the accuracy of blood glucose level estimations using the FTIR device in the MIR range. The performance of models in this study has been measured using various methods including MAE, EMSE and $R^2$ score. Additionally, we employed Clarke Error Grid Analysis [27] and Parkes Error Grid ([28] and [29]) Analysis to further validate the clinical accuracy and utility of our predictive models. We have used LLOCV(Leave one out cross-validation) in order to have all samples both as validation and training in virtue of a low number of samples based on the study goals.

Using LLOC, I identified the hyperparameters for each model, with their ranges displayed in Table 1. Additionally, I have included the ranges for both TBD and ADPD in Table.

To validate the clinical relevance of our models, we applied Clarke Error Grid Analysis and Parkes Error Grid Analysis.

Figure 3 demonstrates the results for both models and the two innovative approaches. As shown, the performance of both methods is approximately superior for the Support Vector Regression model, which is to be expected given its higher complexity. Additionally, Hyperparameters used for both models are detailed in Table 1.

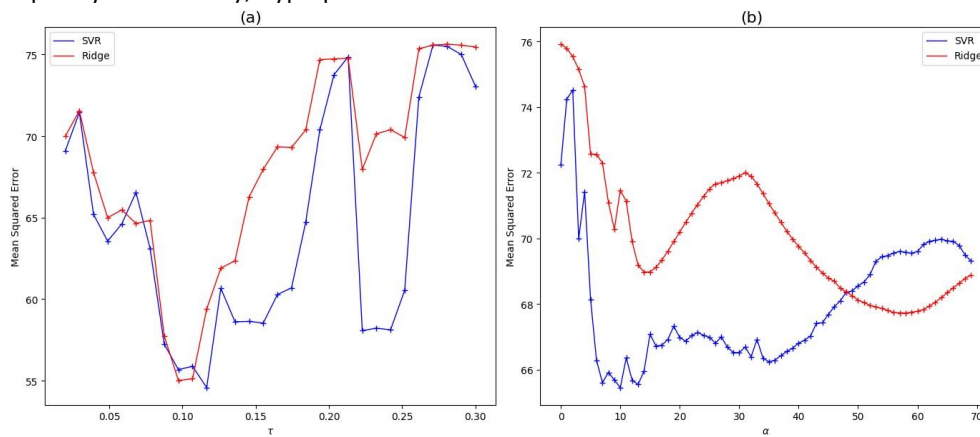

Figure 3: (a): The MSE for both Ridge and SVR in TBD Model (b): The MSE for both Ridge and SVR in ADPD Model

Moreover, Figure 3 shows the results for for the TBD approach, both Ridge and SVR follow similar trends, which suggests a high level of consistency in their performance. This consistency indicates that the TBD approach is stable and effective across different algorithms.

To assess the accuracy of our presented models, we have analyzed four feature selection methods: Base, Derivative, Threshold-Based Derivative (TBD), and Adaptive Derivative Peak Detection (ADPD), in combination with two regression models, Ridge Regression and Support Vector Machine Regression (SVR).

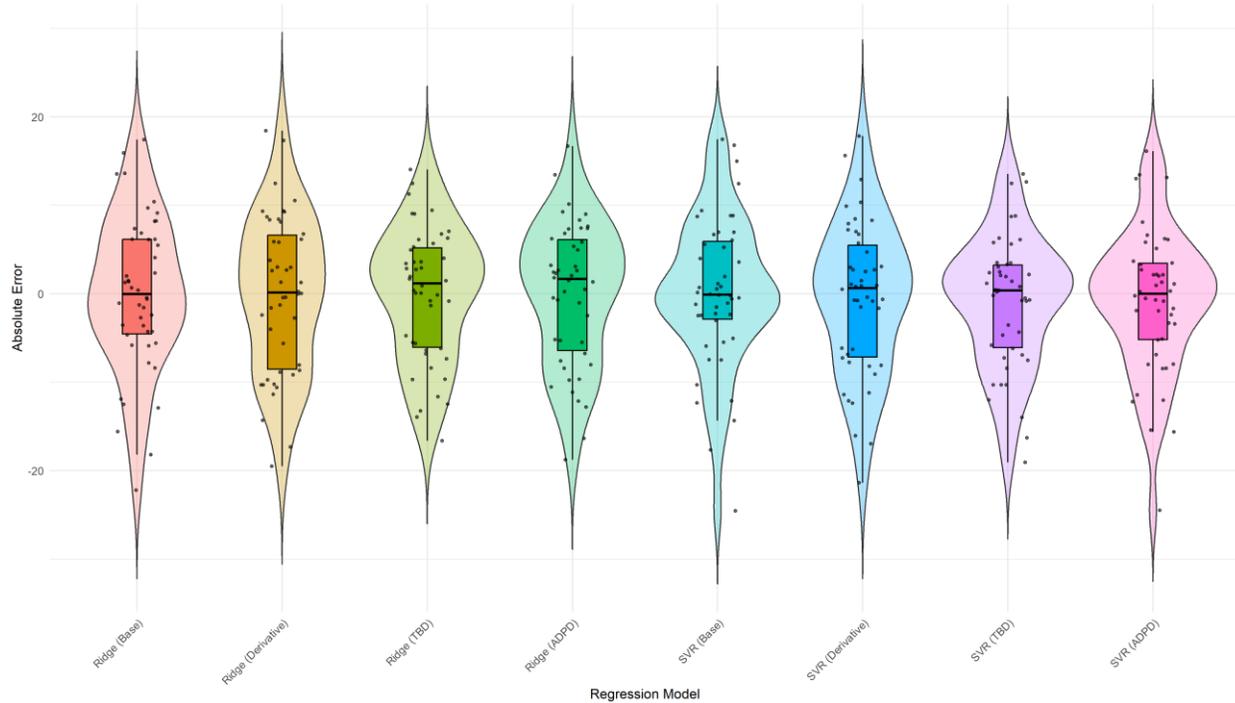

Figure 4: Absolute Error (AE) distribution for Ridge Regression and Support Vector Regression (SVR) using different methods

| Method | Parameter 1 | Tuning Range | Parameter 2 | Tuning Range | Parameter 3 | Tuning Range |
|---|---|---|---|---|---|---|
| TBD | $\tau$ | 0.02–0.3 | - | - | - | - |
| ADPD | $\alpha$ | 0–70 | - | - | - | - |
| Ridge | $\alpha$ | 10–100 | - | - | - | - |
| SVR | $C$ | 0–2 | Kernels | Linear, RBF, Poly | $\epsilon$ | 0.1–0.5 |

Table 1: Tuning ranges of hyperparameters for each model and method (TBD and ADPD) based on LLOCV.

Table 2 provides a comprehensive summary of the performance metrics (MSE, MAE, $R^2$) for both Ridge Regression and SVR across the four methods: Base, Derivative, TBD, and ADPD. The TBD method consistently yields the best performance across both regression models, with the lowest MSE and MAE and the highest $R^2$ values. This suggests that the TBD method effectively captures the relevant features, making more accurate predictions. SVR's performance, particularly in terms of MSE and MAE, aligns with expectations. Its ability to model more complex, nonlinear relationships between features and target variables is evident, as it outperforms Ridge Regression in aspects.

To emphasize the TBD strength further, we have analyzed the Absolye values of each point. As shown in Figure Figure 4, the findings display the Absolute Error distribution using LOOCV.

The TBD approach consistently outperforms the other methods. Both SVR and Ridge Regression utilizing the TBD method showed considerably reduced AE values. The violin plots for TBD indicate a narrower and more focused range of AE values, which reflects better stability and consistency in predictions across different samples. In contrast, the Base and Derivative methods exhibited a wider range of AE values, indicating increased prediction variability.

For further analysis of the two presented pre-processing methods, we observed the effect of TBD and ADPD in SVR, which has demonstrated the best results based on Table 2 compared to Ridge regression, and we compared each point with base pre-processing with presented methods using a Real vs. Predicted plot as it has been exhibited in Figure 5. For TBD and ADPD, the results improved and brought the points closer to the $x = y$ line. However, in both methods, the base model is doing better at some points, and ultimately, some points are getting far from line $y = x$. However, as a whole, the methods enhanced the model's accuracy, and most of the points are getting closer than getting farther from the line, which is why this visualization is also aligned with former results as it should be. Also, as mentioned, it is visually apparent that TBD is doing better compared to ADPD in most of the points, and as a whole, which is aligned with the results based on Table 2.

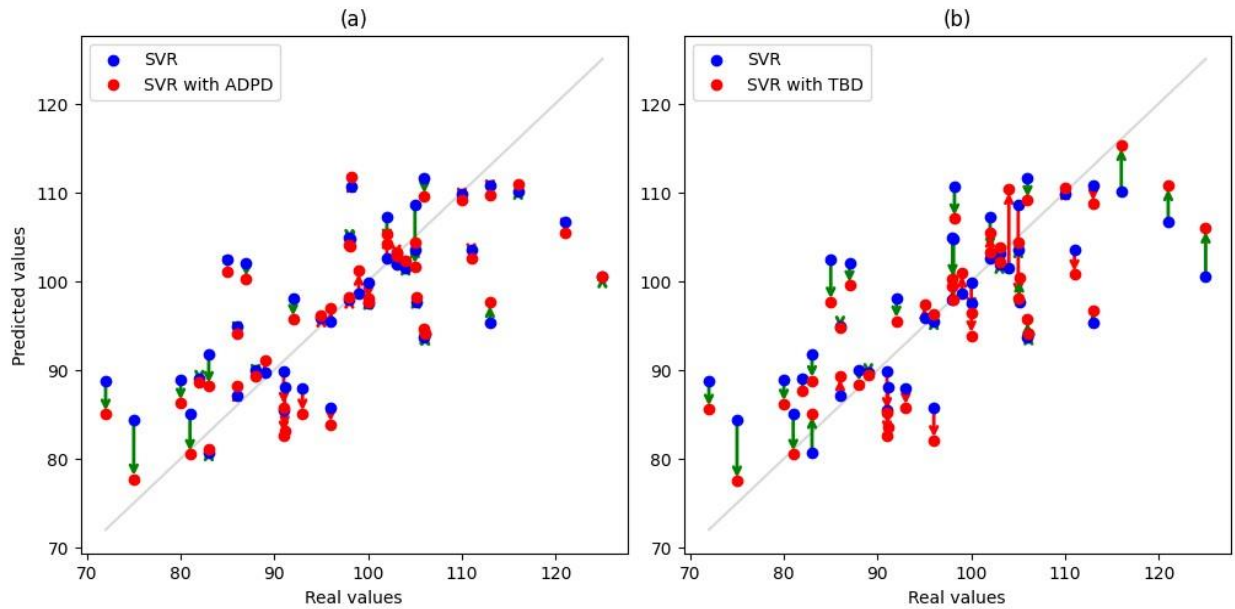

Figure 5: (a): Real vs. Predicted Outcomes for SVR, which shows how ADPD can increase our accuracy (b): Real vs. Predicted Outcomes for SVR, which shows how TBD can increase our accuracy

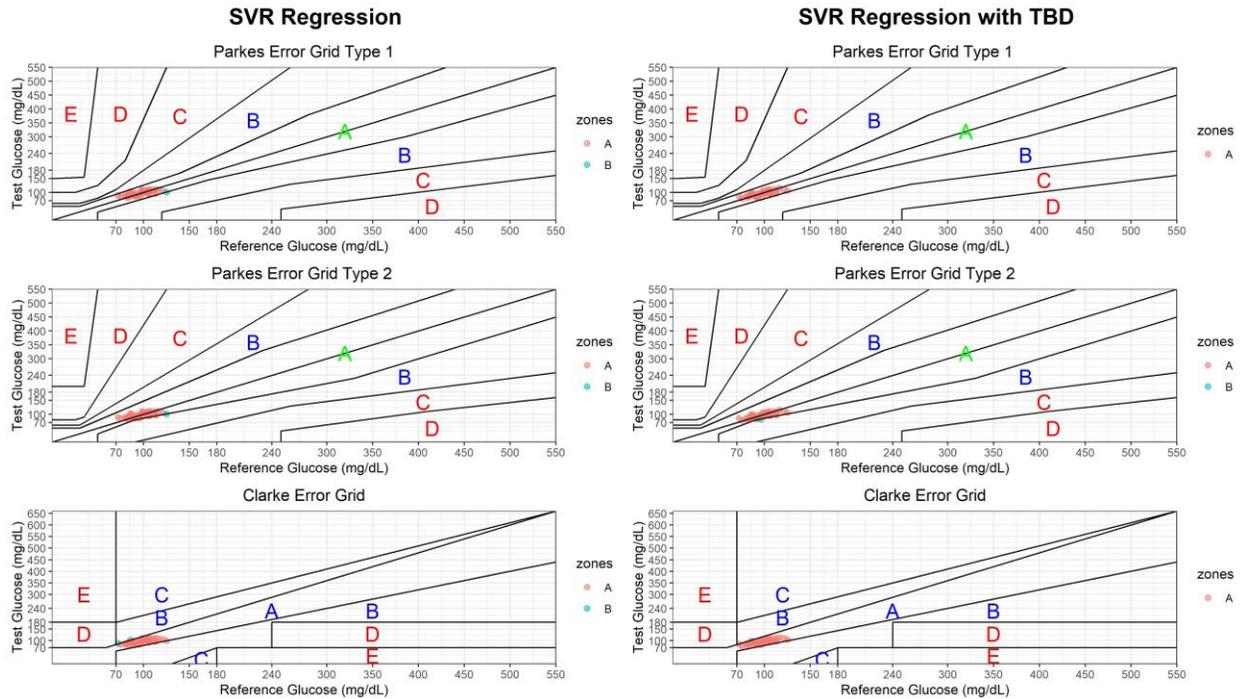

Figure 6: Completion of Base and TBD methods for SVR using Parkes and Clarke Error Grids, emphasizing the improved accuracy of TBD in most clinical zones

As mentioned before, to assess the clinical accuracy of our models, we applied both Clarke Error Grid Analysis (CEG) and Parkes Error Grid Analysis (PEG), standard tools for evaluating the performance of glucose prediction models in clinical settings. The whole result for each method and both models has been shown in Tabel 3, all the points for all processes and models were in the zone A or B, and the Tabel 3 shows the number of points in the zone B. Generally, the TBD and ADPD were superior compared to two others. However, there are some exceptions; for example, the derivative method in the SVR model has no point in both PEGs in zone A. However, the ADPD has one point in zone B for both PEG type 1 and type 2, or compared to TBD in SVR; the TBD has one point in zone B. However, in CEG, the ADPD and TBD always do better than the other models. Also, for further investigation, we chose SVR with TBD as the best method, plotted ite Error Grids, and compared it with basic SVR in Figure 6. The TBD is superior in all three Error Grids, except PEG type 2, which both have the same accuracy. As a result, the innovative methods have also shown more strength in clinical measurements.

| Method | Ridge | | | | SVR | | | |
|---|---|---|---|---|---|---|---|---|
| | Base | Derivative | TBD | ADPD | Base | Derivative | TBD | ADPD |
| MSE | 75.9 | 77.33 | 55.2 | 67.79 | 72.24 | 73.94 | 54.53 | 65.45 |
| MAE | 6.85 | 7.17 | 6.05 | 6.61 | 6.26 | 6.73 | 5.63 | 6.03 |
| $R_2$ | 0.44 | 0.43 | 0.596 | 0.533 | 0.47 | 0.46 | 0.6 | 0.521 |

Table 2: The result of each regression model on the dataset, as we can see approximately SVR has done better, and in each model, TBD has the best accuracy

| Method | Ridge | | | | SVR | | | |
|---|---|---|---|---|---|---|---|---|
| | Base | Derivative | TBD | ADPD | Base | Derivative | TBD | ADPD |
| PEG Type 1 | 1 | 0 | 0 | 0 | 1 | 0 | 0 | 1 |

| PEG Type 2 | 1 | 0 | 1 | 0 | 1 | 0 | 1 | 1 |
|---|---|---|---|---|---|---|---|---|
| CEG | 2 | 2 | 0 | 1 | 2 | 1 | 0 | 0 |

Table 3: Number of points outside zone A in Parkesand Clarke Error Grid for Ridge and SVR models using different methods

In this study, we have introduced two novel preprocessing methods, including Threshold-Based Derivative (TBD) and Adaptive Derivative Peak Detection (ADPD), which enhanced the learning accuracy of blood glucose estimation using MIR spectroscopy data of prior models. The offered methods effectively integrated absorbance spectra and their derivatives, balancing peaks in absorbance data with improved spectral differentiation. Results using Ridge Regression, Support Vector Machine Regression (SVR), and Leave-One-Out Cross-Validation (LOOCV) demonstrated significant improvements in predictive accuracy. The TBD method achieved superior performance metrics for both models, including lower Mean Squared Error (MSE) and Mean Absolute Error (MAE). Additionally, it increased the $r^2$ score by around 36% in Ridge Regression and approximately 27% in SVR. However, these values for ADPD were in order 10% and 24%. Additionally, in all methods, the SVR has superior accuracy compared to Ridge Regression, as it is a more complicated model.

Also, these two models showed better clinical accuracy, as validated by Clarke and Parkes Error Grid analyses. Despite these promising mentioned outcomes, our study is limited by the small dataset size and its focus on the MIR spectral region. Future research will aim to address these limitations by expanding the dataset to include a larger and more diverse population for better generalizability, applying the proposed methods to additional infrared regions, such as the Near-Infrared spectrum.

## References

[1] Pao Lin. Mid-infrared photonic chip for label-free glucose sensing. page JW3A.11, 01 2018.

[2] Jiming Sa, Yuyan Song, Hanwen Gu, and Zhushanying Zhang. Mid-infrared spectroscopy with an effective variable selection method based on mpa for glucose detection. *Chemometrics and Intelligent Laboratory Systems*, 233:104731, 2023.

[3] Noor Nazurah Mohd Yatim, Zainiharyati Mohd Zain, Mohd Zuli Jaafar, Zalhan Md Yusof, Abdur Rehman Laili, Muhammad Hafiz Laili, and Mohd Hafizulfika Hisham. Noninvasive glucose level determination using diffuse reflectance near infrared spectroscopy and chemometrics analysis based on in vitro sample and human skin. pages 30–35, 2014.

[4] Prashant Bhandare, Yitzhak Mendelson, Robert A. Peura, Günther Janatsch, Jürgen D. Kruse-Jarres, Ralf Marbach, and H. Michael Heise. Multivariate determination of glucose in whole blood using partial least-squares and artificial neural networks based on mid-infrared spectroscopy. *Appl. Spectrosc.*, 47(8):1214–1221, Aug 1993.

[5] Guang Han, Siqi Chen, Xiaoyan Wang, Jinhai Wang, Huiquan Wang, and Zhe Zhao. Noninvasive blood glucose sensing by near-infrared spectroscopy based on plsr combines sae deep neural network approach. *Infrared Physics Technology*, 113:103620, 2021.

[6] Bitewulign Kassa Mekonnen, Webb Yang, Tung-Han Hsieh, Shien-Kuei Liaw, and Fu-Liang Yang. Accurate prediction of glucose concentration and identification of major contributing features from hardly distinguishable near-infrared spectroscopy. *Biomedical Signal Processing and Control*, 59:101923, 2020.

[7] Yang Chen, Jiang Liu, Zhenni Pan, and Shirgeru Shimamoto. Non-invasive blood glucose measurement based on mid-infrared spectroscopy. pages 1–5, 2020.

[8] Douglas Carvalho Caixeta, Cassio Lima, Yun Xu, Marco Guevara-Vega, Foued Salmen Espindola, Royston Goodacre, Denise Maria Zezell, and Robinson Sabino-Silva. Monitoring glucose levels in urine using ftir spectroscopy combined with univariate and multivariate statistical methods. *Spectrochimica Acta Part A: Molecular and Biomolecular Spectroscopy*, 290:122259, 2023.

[9] Thomas G. Mayerhöfer, Harald Mutschke, and Jürgen Popp. Employing theories far beyond their limits—the case of the (boguer-) beer–lambert law. *ChemPhysChem*, 17(13):1948–1955, 2016.

[10] Xianchun Shen, Shubin Ye, Liang Xu, Rong Hu, Ling Jin, Hanyang Xu, Jianguo Liu, and Wenqing Liu. Study on baseline correction methods for the fourier transform infrared spectra with different signal-to-noise ratios. *Applied Optics*, 57(20):5794–5799, 2018.

[11] Abraham Savitzky and M. J. E. Golay. Smoothing and differentiation of data by simplified least squares procedures. *Analytical Chemistry*, 36(8):1627–1639, 1964.

[12] Jian Zhang and Abdul M. Mouazen. Fractional-order savitzky–golay filter for pre-treatment of on-line vis–nir spectra to predict phosphorus in soil. *Infrared Physics Technology*, 131:104720, 2023.

[13] Abdul Rohman and Y. B. Che Man. Determination of extra virgin olive oil in quaternary mixture using ftir spectroscopy and multivariate calibration. *Journal of Spectroscopy*, 26(3):471376, 2011.

[14] Nur Cebi, Osman Taylan, Mona Abusurrah, and Osman Sagdic. Detection of orange essential oil, isopropyl myristate, and benzyl alcohol in lemon essential oil by ftir spectroscopy combined with chemometrics. *Foods*, 10(1), 2021.

[15] Zanyar Movasaghi Shazza Rehman Abdullah Chandra Sekhar Talari, Marcela A. Garcia Martinez and Ihtesham Ur Rehman. Advances in fourier transform infrared (ftir) spectroscopy of biological tissues. *Applied Spectroscopy Reviews*, 52(5):456–506, 2017.

[16] Abdulrahman Aloraynan, Shazzad Rassel, Chao Xu, and Dayan Ban. A single wavelength mid-infrared photoacoustic spectroscopy for noninvasive glucose detection using machine learning. *Biosensors*, 12(3), 2022.

[17] Luka Jurjevic, Mateo Gašparović, Anita Simic Milas, and Ivan Balenović. Impact of uas image orientation on´ accuracy of forest inventory attributes. *Remote Sensing*, 12(3), 2020.

[18] Jia Song, Lin-na Du, Hong-bin Wang, Jia-hui Lu, and Wei Han. Application of near infrared spectroscopy combined with partial least squares in quantitative analysis of polysaccharide in irpex lacteus fr. mycelia. In *2010 International Conference on Artificial Intelligence and Computational Intelligence*, volume 3, pages 311–314, 2010.

[19] U. Thissen, M. Pepers, B. Üstün, W.J. Melssen, and L.M.C. Buydens. Comparing support vector machines to pls for spectral regression applications. *Chemometrics and Intelligent Laboratory Systems*, 73(2):169–179, 2004.

[20] Cort J. Willmott and Kenji Matsuura. Advantages of the mean absolute error (mae) over the root mean square error (rmse) in assessing average model performance. *Climate Research*, 30(1):79–82, 2005.

[21] Michael A. Babyak. What you see may not be what you get: a brief, nontechnical introduction to overfitting in regression-type models. *Psychosomatic Medicine*, 66(3):411–421, May-Jun 2004.

[22] Arthur E Hoerl and Robert W Kennard. Ridge regression: Biased estimation for nonorthogonal problems. *Technometrics*, 12(1):55–67, 1970.

[23] U. Thissen, M. Pepers, B. Üstün, W.J. Melssen, and L.M.C. Buydens. Comparing support vector machines to pls for spectral regression applications. *Chemometrics and Intelligent Laboratory Systems*, 73(2):169–179, 2004.

[24] Vladimir Vapnik, Steven Golowich, and Alex Smola. Support vector method for function approximation, regression estimation and signal processing. 9, 1996.

[25] Harris Drucker, Christopher JC Burges, Linda Kaufman, Alexander Smola, and Vladimir Vapnik. Support vector regression machines. In *Advances in neural information processing systems*, volume 9, pages 155–161, 1997.

[26] Retno Endah Masithoh, Heru Zaki Amanah, Woo-Sik Yoon, et al. Determination of protein and glucose of tuber and root flours using nir and mir spectroscopy. *Infrared Physics & Technology*, 113:103577, 2020.

[27] W. L. Clarke, D. Cox, L. A. Gonder-Frederick, W. Carter, and S. L. Pohl. Evaluating clinical accuracy of systems for self-monitoring of blood glucose. *Diabetes Care*, 10(5):622–628, Sep-Oct 1987. Comparative Study, Journal Article, Research Support, U.S. Gov't, P.H.S.

[28] J. L. Parkes, S. L. Slatin, S. Pardo, and B. H. Ginsberg. A new consensus error grid to evaluate the clinical significance of inaccuracies in the measurement of blood glucose. *Diabetes Care*, 23(8):1143–1148, Aug 2000. Clinical Trial, Comparative Study, Journal Article.

[29] Andreas Pfützner, David C Klonoff, Scott Pardo, and Joan L Parkes. Technical aspects of the parkes error grid. *Journal of Diabetes Science and Technology*, 7(5):1275–1281, Sep 1 2013. © 2013 Diabetes Technology Society.